\documentclass[twocolumn,pra,aps]{revtex4}
\draft
\newcommand{\beeq}{\begin{equation}}
\newcommand{\eneq}{\end{equation}}
\newcommand{\beeqar}{\begin{eqnarray}}
\newcommand{\eneqar}{\end{eqnarray}}
\begin{document}
\title{Critique of Protective Measurements}
\author{N.D. Hari Dass$^1$\cite{email1} and 
Tabish Qureshi$^2$\cite{email2} }
\address{$^1$ Institute of Mathematical Sciences, Taramani, 
Madras-600113, INDIA.}
\address{$^2$ Material Science Division,
Indira Gandhi Center for Atomic Research, Kalpakkam, INDIA.}

\begin{abstract}
Recently proposed idea of ``protective'' measurement of a quantum state
is critically examined, and generalized.  Earlier criticisms of the
idea are discussed and their relevance to the proposal assessed.
Several constraints on measuring apparatus required by ``protective''
measurements are discussed, with emphasis on how they may restrict
their experimental feasibility. Though ``protective'' measurements
result in an unchanged system state and a shift of the pointer
proportional to the expectation value of the measured observable in the
system state, the actual reading of the pointer position gives rise to
several subtleties. We propose several schemes for reading pointer
position, both when the apparatus is treated as a classical system as
well as when its quantum aspects are taken into account, that address
these issues. The tiny entanglement which is always present due to
deviation from extreme adiabaticity in realistic situations is argued
to be the weakest aspect of the proposal. 
Because of this, one can never perform a protective measurement on a
single quantum system with absolute certainty.  This clearly
precludes an ontological status for the wave function.  
Several other conceptual issues are also discussed.

\end{abstract}

\maketitle

\section{Introduction}

Quantum mechanics is a theory which has been tremendously successful in
explaining how the physical world works, but its measurement aspects
have been plagued with interpretational problems since its inception.
The general credo is that the value of a real physical observable,
described by a Hermitian operator, has meaning only when the system is
in its eigenstate, i.e.,

\begin{equation}
	A|a_i\rangle = a_i|a_i\rangle,
\end{equation}
where $a_i$ is the eigenvalue of $A$ corresponding to the eigenstate
$|a_i\rangle$. Furthermore, if the system is in a state $|n\rangle$
which is not an eigenstate of $A$, a measurement of $A$ can yield as a
result any of the eigenvalues of $A$ while ``collapsing'' $|n\rangle$
to $|a_i\rangle$ at the same time. Thus the outcome of a single
measurement on a single quantum system can not be assigned any
significance. As a corollary, the state of a single quantum system can
not also be attributed any objective significance. The statistical
interpretation, originating in the early works of Einstein \cite{ein}, can be
considered the ``optimal way out'' for this strange aspect of quantum
phenomena. According to this, if $|n\rangle$ has the  (unique) expansion

\beeq
|n\rangle = \sum_i c_i|a_i\rangle
\eneq
the outcome of a large number of measurements of A  on an ensemble of
identically prepared states are $a_i$ with probability $|c_i|^2$ and
the ``expectation value'' of $A$ in $|n\rangle$ is construed as the
ensemble average $\sum_i|c_i|^2a_i$. The eigenvalue condition (1) can
be interpreted as a sort of consistency condition for this
interpretation.  Clearly any other state $|\tilde n\rangle = \sum
\tilde c_i|a_i\rangle$ with $\tilde c_i=e^{i\phi_i}c_i$ will also yield
an identical distribution of $a_i$ as $|n\rangle$ in an ensemble
measurement of $A$.  To determine $|n\rangle$, therefore, many ensemble
measurements have to be carried out with different observables. The
number of such independent ensemble measurements needed to determine
the original state is dictated by the ``size'' of the density matrix
which is the number of independent parameters needed to specify the
density matrix.

Apart from granting only an ``epistemological'' meaning to the quantum
state (wave function), this interpretation leads to a notion of reality
fundamentally different from that in classical mechanics. It also puts
observation or measurements on a totally different footing than in
classical mechanics  (as John Wheeler has succinctly put it, ``no
phenomenon is a phenomenon until it is an observed phenomenon''). At
the same time, the notion of ``collapse'' or the ``projection
postulate'' as enunciated by von Neumann \cite{NEUMANN} leads to its
own set of conceptual difficulties. As the density matrix of a pure
state ($tr\rho = 1=tr\rho^2$) turns into that of a mixed state ($tr\rho
=1, tr\rho^2 <1$) after the ensemble measurement, something that can
never be achieved through an unitary evolution, it appears as if new
elements have to be introduced into the theoretical framework to
accommodate the measurement process. This in a nutshell is the
``measurement problem'' of quantum theory. Proposals to ``solve'' this
fantastic situation are even more fantastic like the Everett many
worlds interpretation \cite{EVERETT} or the GRW proposal \cite{GRW}.
As there are no feasible means of experimentally testing these at the
moment, they remain as merely matters of individual taste.

For a single quantum state, the situation is even more complex. When
the state is {\it a priori} unknown, measurement of any observable is
generically not going to be an eigenstate measurement. Consequently,
after the measurement, the state of the system will change in an
uncontrollable manner. Any number of subsequent measurements are not
going to give information about the original state i.e the average
values of the outcome of repeated measurements have no bearing on the
expectation value of the observable in the original state (for an
interesting twist to this see section IV-D). 
Of course, the expectation value
of any observable $A$ in an a priori known state $|n\rangle$ can always
be calculated. In such a situation  one can also come up with schemes to
to perform a "measurement" of the expectation value as well as the associated 
variance by either using so called reversible measurements \cite{royer}, 
or by avoiding entanglement. 
But one doesn't gain any new information about the system. Even the 
a priori known wavefunction is verified only in a statistical sense. 
In fact one is only performing an ensemble measurement in disguise.
Thus neither the generic ( as opposed to a priori unknown ) state of 
a single quantum system nor expectation
values of observables in it can be given any meaning. The standard
lore, therefore, denies any ``reality'' or ``ontological" meaning to
the wave function.

Therefore, the recent proposal by Aharonov, Anandan and Vaidman (AAV)
\cite{AV,AAV1,AAV2,metastable} of a scheme involving adiabatic
measurements, which they have called ``protective'' measurements,
wherein they have claimed the possibility of measurement of $\langle
A\rangle$ in the state $|n\rangle$ of a single quantum system for any
observable $A$, without disturbing $|n\rangle$, has indeed raised
surprise and scepticism among many
\cite{SCHWINGER,UNRUH,ROVELLI,GHOSE,DU,ALTER1,ALTER2,SAM}.  This
proposal is remarkable from many points of view, all of a fundamental
nature, and therefore deserves the most careful scrutiny. AAV claim to
be able to measure $\langle A \rangle_n$ for any $A,|n\rangle$ whereas
we saw that the standard lore does not allow it even if one is willing
to uncontrollably disturb $|n\rangle$. Even more remarkably, they claim
to be able to do so without disturbing the system at all. This allows
for these protective measurements to be repeated with sufficiently many
observables to completely determine the state modulo an overall phase.
Here again, the number of different observables to be protectively
measured in order to determine the state of the system is governed by
the number of independent parameters in the density matrix. Thus their
proposal, as stressed by them, allows for an ``ontological'' meaning to
the wave function of a single system.

AAV have made many proposals to realize such protective measurements
which can be broadly split into two categories: i) a Quantum Zeno type
measurements made on an {\em a priori known} state of the single system
and ii) an adiabatic measurement made on {\em an a priori unknown}
state of the system which, however, is {\em known} to be a
non-degenerate eigenstate of {\em an a priori unknown Hamiltonian}.
Here we restrict our attention to only the second category which we
feel is the more interesting one.  A number of criticisms of this
proposal have appeared subsequently
\cite{SCHWINGER,UNRUH,ROVELLI,GHOSE,DU,ALTER1,ALTER2,SAM}.  In this
paper we critically review and assess the original proposal as well as
the criticisms. We also extend the scope and generality of both.

The paper is organised as follows: in section II we present 
the idea of protective measurements in a rigorous way, and then go on to 
generalize it. We also discuss a few examples which highlight some subtle 
points regarding the original AAV proposal. In section III, we critically 
analyze various criticisms of the original AAV proposal and assess their 
relevance to the issue. In section IV we
discuss the very important issue of spreading of the pointer position and
suggest some ways to circumvent the problem. In section V we make detailed
remarks on the restrictions imposed on the measuring apparatus by protective
measurements, and the feasibility of practical implementation of the idea. We
also discuss the relevance of protective measurement to the issue of the
"reality" of the wave-function. Finally, in section VI we summarise the
main results of the present investigation.

A more rigorous derivation as well as generalisation of the original AAV
proposal(sec II), a discussion of the relevance of the degeneracy of the total
( system and apparatus ) Hamiltonian with examples ( sec IIB ), a careful treatment of the effects of switching on/off of the apparatus-system interaction ( sec IIC), an unambiguous
rephrasing of the AAV spin-1/2 example ( sec IIE ) are new features of this 
paper designed to bring greater clarity to the discussion. Secs III-V
are totally new contributions.

\section{Protective measurement}

Let us first consider a conventional measurement. Let $Q_S$ be an
operator, corresponding to the observable of the system we wish to
measure, and let it interact with an appropriate apparatus( in what follows,
we shall use the notion of an apparatus to indicate a quantum system to which
full information about the system can be transferred) through
an interaction

\begin{equation}
        H_I = g(t) Q_A Q_S,
\end{equation} 
where $Q_A$ is an observable of the apparatus, and $g(t)$ is the strength
of the interaction normalized such that $\int dt g(t)=1$. The interaction is
nonzero only in the short interval $[0,\tau]$. Let the system be in an 
initial state $|\nu\rangle$
which is not necessarily an eigenstate of $Q_S$, and the apparatus be
in a state $|\phi(r_0)\rangle$, which is a wave packet of eigenstates
of the operator $R_A$ conjugate to $Q_A$, centered at the eigenvalue
$r_0$. The interaction $H_I$ is of short duration, and assumed to be so
strong that the effect of the free Hamiltonians of the apparatus and
the system can be neglected. Then the combined wave function of the
system and the apparatus at the end of the interaction can be written as

\begin{equation}
        |\psi(\tau)\rangle = e^{-{i\over\hbar} Q_A Q_S } |\nu\rangle
|\phi(r_0)\rangle.
\end{equation}
If we expand $|\nu\rangle$ in the eigenstates of $Q_S$,  $|s_i\rangle$, we
obtain
\begin{equation}
|\psi(\tau)\rangle = \sum_{i} e^{-{i\over\hbar} Q_A s_i } c_i |s_i\rangle
|\phi(r_0)\rangle,
\end{equation}
where $s_i$ are the eigenvalues of $Q_S$ and $c_i$ are the expansion
coefficients. The exponential term shifts the center of the wave packet by 
$ s_i $:
\begin{equation}
|\psi(\tau)\rangle = \sum_{i} c_i |s_i\rangle |\phi(r_0+ s_i )\rangle.
\end{equation}
This is an entangled state, where the position of the wave packet gets
correlated with the eigenstates $|s_i\rangle$. Detecting the center of the 
wave packet at $r_0+s_i$ 
will throw the system into the eigenstate $|s_i\rangle$. 

Protective measurements, on the other hand,  make use of the opposite
limit where the interaction of the system with the apparatus is {\it
weak} and {\em adiabatic}.  Here the system is assumed to be in a
non-degenerate eigenstate of its Hamiltonian, and the interaction being
weak and adiabatic, we cannot neglect the free Hamiltonians. Let the
Hamiltonian of the combined system be

\begin{equation}
H(t) = H_A + H_S + g(t)Q_A Q_S, \label{H_full}
\end{equation}
where $H_A$ and $H_S$ are the Hamiltonians of the apparatus and the
system, respectively. The coupling $g(t)$ acts for a long time $T$ and goes
to zero smoothly before and after the interaction.  It is also
normalized as $\int_0^T dt g(t) = 1$. Therefore, $g(t) \approx 1/T$ is
small and constant for the  most part. If $|t=0\rangle$ is the state vector
of the combined apparatus-system just before the measurement process
begins,  the state vector after T is given by

\begin{equation}
|t=T\rangle = {\cal T} e^{-{i\over\hbar}\int_0^T H(\tau) d\tau} |t=0\rangle,
\label{psiT}
\end{equation}
where ${\cal T}$ is the time ordering operator. We divide the interval
$[0,T]$ into $N$ equal intervals $\Delta T$, so that $\Delta T = T/N$, and
because the full Hamiltonian commutes with itself at different times during
$[0,T]$,
we can write eqn(\ref{psiT}) as

\begin{equation}
|t=T\rangle = \left(exp[-{i\Delta T\over\hbar}(H_A + H_S +
{1\over T}Q_A Q_S)]\right)^N |t=0\rangle .
\end{equation}

Let us now examine the case when $Q_A$ commutes with the free
Hamiltonian of the apparatus, i.e., $[Q_A,H_A]=0$, so that we can have
eigenstates $|a_i\rangle$ such that $Q_A |a_i\rangle = a_i |a_i\rangle$
and $H_A |a_i\rangle = E_i^a |a_i\rangle$. Choudhury, Dasgupta and
Datta \cite{DU} consider only two cases: one where $[Q_A,H_A]=0$ and
$[Q_S,H_S]=0$, second where $[Q_A,H_A]\neq 0$ and $[Q_S,H_S]\neq 0$.
Thus they put an additional restriction that $Q_A$ and $Q_S$ either
commute or do not commute with the unperturbed Hamiltonian, together,
and miss the important case where $[Q_A,H_A]=0$ and $[Q_S,H_S]\neq 0$.
Now $|a_i\rangle$ are also the exact eigenstates of the instantaneous
Hamiltonian $H(t)$, in the apparatus subspace. So, the exact
instantaneous eigenstates can be written in a factorized form
$|a_i\rangle \overline{|\mu\rangle}$ where $\overline{|\mu\rangle}$ are
system states which depend on the eigenvalue of $Q_A$, {\it i.e.}, they
are the eigenstates of ${1\over T}a_iQ_S + H_S$. Let us assume the
initial state to be a direct product of a non-degenerate eigenstate of
$H_S$, $|\nu\rangle$, and $|\phi(r_0)\rangle$:

\begin{equation}
|t=0\rangle = |\nu\rangle |\phi(r_0)\rangle .
\end{equation}
Introducing complete set of  exact eigenstates in the above equation,
the wave function at a time $T$ can now be written as

\begin{equation}
|t=T\rangle = \sum_{i,\mu} e^{{i\over\hbar}E(a_i,\mu) N\Delta T}
|a_i\rangle \overline{|\mu\rangle}\overline{\langle\mu|} 
|\nu\rangle \langle a_i| |\phi(r_0)\rangle, \label{psiT1} 
\end{equation}
where the exact instantaneous eigenvalues $E(a_i,\mu)$ can be written as

\begin{equation}
E(a_i,\mu) = E_i^a + {1\over T} \overline{\langle\mu|}Q_S\overline{|\mu\rangle}
a_i + \overline{\langle\mu|}H_S\overline{|\mu\rangle}.
\end{equation}

Till here the treatment is exact, except for ignoring the switching on and
switching off times to begin with. We justify ignoring these in sec IIC. 
It should be kept in mind that the
expectation value $\langle Q_S\rangle _{\overline{\mu}}$ depends on the eigenvalue
$a_i$ of $Q_A$. The sum over $\mu$ in (\ref{psiT1}) makes it appear as
if the state is entangled. But the important point to notice is that
the basis $\overline{|\mu\rangle}$ can be made to be {\em arbitrarily}
close to the original basis, as the interaction is assumed to be weak,
so that $\overline{|\mu\rangle} = |\mu\rangle + {\cal O}(1/T) + ...$.
In the large $T$ limit, one can assume the states to be unperturbed,
and retain only terms of $O(1/T)$ in the energy (this is necessary as
$E(a_i,\mu)$ is multiplied by T in eqn(11)), which amounts to using
first order perturbation theory. This yields eigenvalues of the form

\begin{equation}
E(a_i,\mu) = E_i^a + {1\over T} \langle\mu|Q_S|\mu\rangle a_i + 
\langle\mu|H_S|\mu\rangle + O(1/T^2).
\end{equation}
In addition to this, the sum over $\mu$ disappears and only the
term where $\mu =\nu$ survives.  Thus, we can write
the apparatus part of the exponent again in the operator form

\begin{equation}
|t=T\rangle \approx e^{-{i\over\hbar}H_A T-{i\over\hbar}
Q_A\langle Q_S\rangle _\nu -{i\over\hbar}\langle H_S\rangle _\nu T}
|\nu\rangle |\phi(r_0)\rangle .
\end{equation}

Now, it is easy to see that the second term in the exponent will shift
the center of the wave packet $|\phi(r_0)\rangle$ by an amount $
 \langle\nu|Q_S|\nu\rangle$:

\begin{equation}
|\psi(T)\rangle = e^{-{i\over\hbar}H_A T-{i\over\hbar}\nu T}
|\nu\rangle |\phi(r_0+\langle Q_S\rangle _\nu)\rangle .\nonumber\\
\end{equation}
This shows that at the end of the interaction, the center of the
wave packet $|\phi(r_0)\rangle$ shifts by $\langle\nu |Q_S|\nu\rangle$.

The idea behind this approximation is that in $\overline{\langle
\mu}|\nu\rangle$ only one term is large and close to unity, and rest of
the terms are very small, of the order $1/T$. Making $T$ very large,
one can make the smaller terms arbitrarily close to zero. Thus, the
state is effectively not entangled, and so the original wave function is
not destroyed during the measurement.  Looking at the position of the
wave packet, one can determine the expectation value  $\langle
Q_S\rangle _\nu $. This, basically, is the essence of the argument for
protective measurements,
although it was not shown with this much rigour in the original proposal.
Further, it has been asserted that one {\it needs} the
condition $[Q_A,H_A]=0$ to obtain a clean protective measurement
\cite{metastable}. In the following we will show that this condition is
not really necessary for a protective measurement, and the idea can be
made quite general.

\subsection{The general case}

We consider again the Hamiltonian in (\ref{H_full}). As we are
interested in examining the possibility of protective measurements in
the most general context,

\beeq
[H_A,Q_A]\ne 0~~~~~~~~~~~~[H_S,Q_S]~\ne 0.
\eneq
T denotes the duration of the adiabatic measurement. If $|t=0\rangle$
is the state vector just before the measurement process begins, the
state vector after T is again given by (\ref{psiT}). Here again, with
$g(t)=1/T$, the Hamiltonian is time-independent and no time-ordering is
needed. In that case

\beeq
|t=T\rangle= e^{iTH}|t=0\rangle ,
\eneq
where 

\beeq
H= H_A+H_S+{Q_AQ_S\over T} \label{HdT}.
\eneq
We start with an initial state satisfying the conditions laid down by
{\cite{AV,AAV1,AAV2}

\beeq
|t=0\rangle  =  |\nu\rangle |\phi\rangle ,
\eneq
where $|\nu\rangle$ is a non-degenerate eigenstate of $H_S$ and
$|\phi\rangle$ is a general state of the apparatus, not necessarily an
eigenstate of $H_A$ (which we shall denote generically by $|a\rangle$).
Then

\beeq
|t=T\rangle  = e^{iHT} |\nu\rangle|\phi\rangle .
\eneq
We further expand $|\phi\rangle$ in the basis $|a\rangle$ and write

\beeq
|t=T\rangle  = e^{iHT}\sum_b d_b |\nu\rangle|b\rangle .
\eneq
Denoting the exact eigenstates of $H$ by $|\Psi_{\mu,a}\rangle$ and the
corresponding eigenvalues by $E(\mu,a)$, we have

\beeq
|t=T\rangle = \sum_b d_b \sum_{\mu,a} e^{iE(\mu,a)T} \langle
\Psi_{\mu,a}|\nu,b\rangle |\Psi_{\mu,a}\rangle .
\label{psiT2}
\eneq

So far no approximations have been made, except of course, for ignoring the 
switching on and switching off times in the beginning (see, however, sec IIC). 
The Hamiltonian $H$ of eqn
(\ref{HdT}) can be thought of as $H_0=H_A+H_S$ perturbed by
${Q_AQ_S\over T}$.  Using the fact that ${Q_AQ_S\over T}$ is a small
perturbation and that the eigenstates of $H_0$ are of the form
$|\nu\rangle|a\rangle$, perturbation theory gives

\beeqar
|\Psi_{\mu,a}\rangle &=& |\mu\rangle|a\rangle +O(1/T)+....\nonumber\\
 E(\mu,a) &=& \mu + E_A(a)+{1\over T}\langle Q_S\rangle_\mu\langle 
Q_A\rangle_a +.... . \label{perturb}
\eneqar

An important qualification needs to be made here. It is important for
eqn(\ref{perturb}) to hold that $|\mu\rangle|a\rangle$ be a {\em
non-degenerate} eigenstate of $H_0=H_A+H_S$, except when the degeneracy arises
solely due to the degeneracy of the eigenstates of $H_A$. Otherwise, even 
in the
limit $T\rightarrow \infty$, the exact eigenstates of $H$ do not
approach $|\mu\rangle|a\rangle$.  We discuss this aspect in more detail
in the following subsection with the help of two illustrative examples.

On substituting 
eqn(\ref{perturb}) in eqn(\ref{psiT2}) and taking the large T limit yields

\beeq
|t=T\rangle  =  \sum_b e^{i(\nu T+E_A(b)T+\langle Q_A\rangle_b\langle 
Q_S\rangle_\nu)} d_b |b\rangle|\nu\rangle .
\eneq
We now introduce the operator

\beeq
Y=\sum_b\langle Q_A\rangle_b|b\rangle\langle b| .
\eneq
It is important to note that the operator $Y$ is {\em a property of the
apparatus alone and does not depend on the system}. In terms of $Y$, the
above eqn can be recast as

\beeq
|t=T\rangle  =  e^{i\nu T} e^{iH_AT} e^{iY\langle Q_S\rangle_s}  
|\phi\rangle|\nu\rangle .
\eneq
If $|\phi\rangle$ of the apparatus is so chosen that it is peaked around a
value $x_0$ of the operator $X$ (the pointer variable) conjugate to $Y$
i.e $[Y,X]=i\hbar$,

\beeq
e^{iX\langle Q_S\rangle_\nu} |\phi(x_0)\rangle = |\phi(x_0+
\langle Q_S\rangle_\nu)\rangle .
\eneq
Thus, modulo the issue of the ``spreading of the pointer position'' by
$H_A$, which is present in any case even in the special case discussed
earlier, the protective measurement of $\langle Q_S\rangle_\nu$
without disturbing $|\nu\rangle$ is a generic possibility.  It should
of course be pointed out that on the one hand it may not always be
possible to physically realize the operator Y, and on the other hand an
operator canonically conjugate to Y need not always exist. For example,
there is no operator canonically conjugate to $X^2$. These and the
restrictions due to degeneracy of $H_0$ may severely restrict the
choice of realistic possibilities.

\subsection{Degeneracy of $H_0$-eigenstates}

As we discussed earlier, in order that eqn(\ref{perturb}) holds, we
require that $|\mu\rangle|a\rangle$ be a {\em non-degenerate}
eigenstate of $H_0=H_A+H_S$.  However, the case where such degeneracy
is due to the degeneracy of eigenstates of $H_A$ alone, is not really a
problem as a suitable basis in the degenerate subspace can be chosen in
terms of which eqn(\ref{perturb}) still holds good. We give two examples to
clarify this aspect.

{\bf i) Two Harmonic Oscillators}\\
Let us consider the situation where both the apparatus and the system are
harmonic oscillators with frequency $\omega$. Thus
\beeqar
H_A&=& P^2/2M+1/2M\omega^2X^2 ,\nonumber\\
H_S&=& p^2/2m+1/2m\omega^2x^2 .
\eneqar
The energy eigenvalues for the eigenstates of this combined system labeled
by $|N,n\rangle=|N\rangle |n\rangle$ are
\beeq
E(N,n)=\hbar\omega(N+n+1). \nonumber\\
\eneq
For example, the state $|0,0\rangle$ is non-degenerate, but the states $|1,0\rangle,|0,1\rangle$
are degenerate. Now consider the adiabatic interaction
\beeq
H_I=g(t)X\cdot x .
\eneq
Let us concentrate on a degenerate subspace in the sum over $(\mu,a)$
in eqn(22).  For illustration, let us choose the subspace with energy
$E(0,1)=E(1,0)$. The unperturbed states are $|1\rangle|0\rangle$ and
$|0\rangle|1\rangle$ respectively. The interaction $H_I$ lifts the
degeneracy and the eigenstates of $H=H_0+H_I$ are

\beeq
|\pm\rangle ={|1,0\rangle\pm|0,1\rangle\over \sqrt 2},
\eneq
with energy eigenvalues $E_{\pm}=2\hbar\omega \pm g\lambda$ where
$\lambda= \langle 0|X|1 \rangle \cdot \langle 1|x|0\rangle$.  Thus if
the initial state were of the type $\sum_Nd_N|N\rangle|0\rangle$, the
contribution in eqn(22) proportional to $d_1$ would be

\beeq
e^{iE_+T}\langle +|1,0\rangle |+\rangle+e^{iE_-T}\langle -|1,0\rangle 
|-\rangle .
\eneq
After some simplifications this reduces to
\beeq
e^{i2\hbar\omega T}\{\cos{g\lambda T}|1,0\rangle+i\sin{g\lambda T}|0,1\rangle\},
\eneq
which in the $T\rightarrow\infty$ limit reduces to
\beeq
e^{i2\hbar\omega T}\{\cos{\lambda }|1,0\rangle+i\sin{\lambda }|0,1\rangle\}.
\eneq
This introduces strong entanglement between the apparatus and system even in the adiabatic limit 
and consequently no protective measurement is possible.

{\bf ii) Harmonic Oscillator Coupled to spin-1/2}\\
Let us consider a spin-1/2 particle(system) coupled to a harmonic 
oscillator (apparatus). The total Hamiltonian is
\beeq
H= P^2/2M+1/2M\omega^2X^2+\mu B_0\sigma_z+gX\vec\sigma\cdot\vec n .
\eneq
With the choice $\mu B_0=1/2\hbar\omega$, we see that the states
$|0\rangle|+\rangle$ and $|1\rangle|-\rangle$ are degenerate. Also, the
interaction Hamiltonian $H_I=gX\vec\sigma\cdot\vec n$ is not diagonal
in this degenerate subspace.  Again, there will be strong entanglement
between the apparatus and system even in the adiabatic limit.

What one learns from these examples is that whenever the eigenstates of
$H_0$ are degenerate in the sense
mentioned above, and when the interaction Hamiltonian $H_I$ is not
diagonal in that degenerate subspace, entanglement between the
apparatus and system can not be avoided even in the adiabatic limit.
These two examples are cases of what could be called ``accidental"
degeneracy of $H_0$.

It is also clear that whenever either $H_A$ or $H_S$ has a continuous
spectrum, $H_0$ generically has degenerate eigenstates. As an example,
consider the situation where $H_A$ has continuous spectrum $a^2$, and
$H_S$ the discrete spectrum $\pm \mu B_0$. Clearly the states
$|a\rangle$ and $|a^\prime\rangle$ are degenerate whenever
${a^\prime}^2=a^2+2\mu B_0$. It is obvious that ${a^\prime}^2\ge 2\mu
B_0$. This is an example of what we call ``generic" degeneracy of
$H_0$.  Protective measurement in such cases is possible only if $H_I$
is diagonal in the degenerate subspace.  In the case when
$[H_A,Q_A]=0$, $H_I$ is indeed diagonal in the respective degenerate
subspace and protective measurement is possible as we saw in section II. When
$[H_A,Q_A]\ne 0$, the situation is more complex.  For $H_I$ to be
diagonal in the degenerate subspace requires $\langle a|Q_A|a^\prime
\rangle=0$ whenever ${a^\prime}^2=a^2+2\mu B_0$ for the example
considered ($\langle a|Q_A|a^\prime\rangle =0$ for {\em all}
$a,a^\prime$ would have meant $[H_A,Q_A]=0$). This already precludes the
prototypical Hamiltonian for Stern-Gerlach experiments:  \beeq
H=P^2/2M+\mu B_0 \sigma_z+\mu B_iX\vec\sigma\cdot\vec n \eneq The only
reason the AAV spin-1/2 example works is because of the assumption
$P^2/2M\simeq 0$. 
We shall see this more clearly in section IIE.

\subsection{Switching on/off of the interaction}

In our treatment so far, we have ignored the possible effects of the
switching on and switching off of the apparatus- system interaction. This may
appear at first to question the use of the adiabatic treatment. However, it
should be borne in mind that the change in the total Hamiltonian during these
periods being $Q_AQ_S/T$ is very small and the switching on and off is really
a gentle process. Therefore it is intuitively clear that no violence has been
committed against the adiabaticity of interactions. Nevertheless, it is
desirable to put this intuitive feeling on a firmer mathematical ground to
make sure nothing subtle has been missed out.

For this purpose let us assume that the interaction is smoothly switched on
during the period $0 \leq t \leq \Delta T $. During this period let the function
$g(t)$ be smooth and bounded by $1/T$ i.e $|g(t)|\leq 1/T$. We can also
arrange for $g(t)$ to be monotonically increasing, but this is not crucial.

Now let us divide the interval $[0,\Delta T] $ into $M$ equal parts of $\tau$ each. The initial Hamiltonian is then $H_0$ and the final Hamiltonian is
$H_0+Q_AQ_S/T$. During the interval labelled by m, the Hamiltonian is
\begin{equation}
H^{(m)} = H_0 + g_m Q_AQ_S
\end{equation}
Let the exact eigenstates and eigenvalues of this Hamiltonian be 
$ |\Psi_{\mu,a}^{(m)}\rangle$,
$E_{\mu,a}^{(m)}$. As the Hamiltonian is {\it time-dependent} now, it is necessaryto use time-ordered products. The state at $t = \Delta T$ is given by
\begin{equation}
|\Delta T\rangle = \prod_m e^{iH_m\tau} |t=0\rangle 
\end{equation}
In a manner analogous to how we obtained eqn (22) we now get

\beeqar
|\Delta T\rangle &=& \sum_b d_b \sum_{\mu_1,\mu_2,..,\mu_M;a_1,a_2,..,a_M}
\nonumber\\
&& e^{i\tau (E^{(1)}_{\mu_1,a_1}+E^{(2)}_{\mu_2,a_2}+...+E^{(M)}_{\mu_M,a_M})}
|\Psi^{(M)}_{\mu_M,a_M}\rangle \nonumber\\
&& \langle \Psi^{(M)}_{\mu_M,a_M)}|\Psi^{(M-1)}_ {\mu_{M-1},a_{M-1}}\rangle.....
\nonumber\\
&& \langle \Psi^{(1)}_{\mu_1,a_1)}|\Psi^{(0)}_ {\mu_{0},a_{0}}\rangle.....
\eneqar
Because the Hamiltonians at adjacent time intervals $(i,i+1)$ differ by 
$(g_i-g_{i+1})Q_AQ_S$ which is again small and bounded by $Q_AQ_S/T$, we
have
\beeqar
\langle \Psi^{(i+1)}_{\mu_{i+1},a_{i+1}}|\Psi^{(i)}_{\mu_i,a_i}\rangle
&=& \delta_{\mu_{i+1},\mu_i}\delta_{a_{i+1},a_i}\nonumber\\
 &+& (g_{i+1}-g_i)({\cal A}+O(1/T))+..
\eneqar
Here ${\cal A}= \langle \nu,b|Q_AQ_S|\nu,b\rangle$ and dots refer to terms higher 
order in $1/T$. Likewise, the energy eigenvalues satisfy
\begin{equation}
E^{(i)}_{\mu_i,a_i} = E_{\nu,b} + g_i {\cal A}
\end{equation}
Combining these eqns and taking the limit $M$-large, one gets
\begin{equation}
|\Delta T\rangle = e^{i(\nu+E_b^A)\Delta T + \int_0^{\Delta T} dt g(t) {\cal A}}d_b|b\rangle|\nu\rangle
\end{equation}
On comparing with eqn(24) it can be seen that the effect of smoothly switching 
on the interaction in the interval $(0,\Delta T)$ can be completely ignored. 
The same applies for the 
interval when the interaction  is smoothly switched off too.

\subsection{An Example with $[H_A,Q_A]=0$}

Let us now consider a specific example embodied by the Hamiltonian

\beeq
H = {P^2\over 2M}+\mu B_0\sigma_z+g(t) \mu B_i P \vec\sigma\cdot \vec n,
\eneq
where $M$ is the mass of the particle with spin whose position acts as
an apparatus, $\mu$ the magnetic moment of the particle, $B_0$ the
homogeneous magnetic field that breaks the degeneracy of $H_S$, $B_i
P\vec n$ a {\em momentum dependent} magnetic field that couples the
apparatus and system degrees of freedom ($\vec \sigma$). Thus in this
example $[H_A,Q_A]=0$ while $[H_S,Q_S]\ne 0$. Further, $\nu = \pm \mu
B_0$ while $E_A(a)=a^2/2M$. We take the initial state to be

\beeq
|t=0\rangle =|\phi(\epsilon,0)\rangle|+\rangle ,
\eneq
where $|\phi(\epsilon,0)\rangle$ is a wave packet of width $\epsilon$
centered at $x=0$. It is clear from the general discussion that in this
case $Y=P$ and that the pointer is the center of the wave packet. In
position representation

\beeq
\langle x|\phi(\epsilon ,0)\rangle  = 
\epsilon^{-1/2}\pi^{-1/4} e^{-{x^2\over 2\epsilon^2}}.
\eneq
We can decompose this wave packet in terms of the plane wave states
(eigenstates of $H_A$) 

\beeq
d(a) = {1\over \sqrt{2\pi}}\int dx e^{-iax} \langle x|\phi(\epsilon,0)\rangle .
\eneq
One obtains

\beeq
d(a) = \pi^{-1/4} \epsilon^{1/2} e^{-{a^2\epsilon^2\over 2}}.
\eneq
Combining these details with eqn(13) one finds that in the case of this
example

\beeq
|t=T\rangle  = e^{i\mu B_0T} e^{i{P^2\over 2M}T} e^{iP\mu
B_i\langle\vec\sigma\cdot\vec n\rangle_+} |+\rangle|\phi(\epsilon,0)\rangle .
\eneq
The operator $e^{iP\mu B_i\langle\vec\sigma\cdot\vec n\rangle_+}$
only shifts the center of the wave packet without changing its width and
$e^{i{P^2\over 2M}T}$ only spreads the wave packet without shifting the
center. Thus we find

\beeq
|t=T\rangle  = e^{iB_0T}|+\rangle|\phi(\epsilon(T),\mu B_i
\langle\vec\sigma\cdot\vec n\rangle)\rangle ,
\eneq
where

\beeq
\epsilon(T)^2= {1\over 2}(\epsilon^2+{T^2\over M^2\epsilon^2})
\eneq
is the standard formula for the spreading of the wave packet. One may
note that the spread in the pointer position in this example is
independent of the system state.}

\subsection{AAV Spin-1/2 Example}

The AAV example of protective measurement on a spin-1/2 state by an
inhomogeneous magnetic field attracted a lot of criticism
\cite{SCHWINGER,UNRUH,ROVELLI,GHOSE,DU,ALTER1,ALTER2,SAM}. We present
here what we think is a better way to look at this example in order to
avoid any confusion. 
We take the inhomogeneous field to be $B_i x \vec
n$. We take $H_A=0$ or equivalently ignore $P^2/2M$. The relevant
Hamiltonian is

\beeq
H = -\mu B_0\vec\sigma\cdot\vec {\tilde n}-\mu g(t) B_i x\vec\sigma\cdot\vec n.
\eneq
As before, $g(t)$ is taken to be ${1\over T}$. {\em It should be noted
that $B_0\vec n$ is an a priori unknown magnetic field}. Consequently
we shall not assume anything about the size of $B_0$. The initial state
is chosen to be

\beeq
|t=0\rangle = e^{ip_0x} |\tilde +\rangle ;~~~~~~~\vec\sigma\cdot\vec{\tilde n}
|\tilde \pm\rangle=\pm|\tilde \pm\rangle .
\eneq
It should be emphasized that this initial state is a priori unknown.
The Hamiltonian of eqn(37) is the Hamiltonian of the spin-1/2 particle
in the effective magnetic field

\beeq
\vec B=B_0\vec{\tilde n}+B_i {x\over T}\vec n ,
\eneq
whose eigenstates are given by
\beeq
H|\pm\rangle=\pm \mu B|\pm\rangle .
\eneq
Consequently, the state at $t=T$ is given by
\beeq
|t=T\rangle=[\cos {\theta\over 2} e^{i\mu BT} |+\rangle+\sin{\theta\over 2} 
e^{-i\mu BT} |-\rangle ,
\eneq
where $\theta$ is the angle between $\vec B$ and $\vec{\tilde n}$. As $T\rightarrow \infty$,
$\theta\rightarrow 0$ and $|+\rangle\rightarrow|\tilde +\rangle$. Also
\beeq
B\rightarrow B_0+B_i{x\over T}\vec n\cdot\vec{\tilde n}.
\eneq
Thus
\beeq
|t=T\rangle  \rightarrow e^{i\mu B_0T} e^{i(p_0+\mu B_i
\vec n\cdot\vec{\tilde n}x)} |\tilde +\rangle .
\eneq
Hence the momentum of the apparatus shifts by $\mu B_i\vec
n\cdot\vec{\tilde n}= \langle\mu B_i\vec\sigma\cdot\vec
n\rangle_{\tilde +}$, while the system remains in the same state to
begin with.

The language used inadvertently by AAV in describing this example has,
in our view, been partly responsible for some of the misunderstandings
about the AAV proposal engendering a class of criticisms in
\cite{SCHWINGER,UNRUH,ROVELLI,GHOSE,DU,ALTER1,ALTER2,SAM}. For example
AAV say, ``$B_0$ is very large compared to the Stern-Gerlach field". This
unnecessarily gives the impression that $B_0$ is a priori known and
consequently $|\tilde +\rangle$ is a priori known too.  A less
confusing way to state this would have been to say that because of
adiabaticity the Stern-Gerlach field $ B_i{x\over T}$ can be made much
smaller than any $B_0$. Likewise, AAV state, ``to see the transition
from the usual Stern-Gerlach case, we may gradually increase $B_0$ from
0".  This too gives the same false impression of $B_0$ being known (and
hence controllable)  a priori.  In fact, while the usual Stern-Gerlach
set up involves an impulsive transition, the modified Stern-Gerlach set
up involves an adiabatic transition.  This can be understood as arising
out of tuning $B_0$ only in a formal way.

\section{Assessing the criticisms}

The proposal of protective measurements drew a lot of criticism on various
counts 
\cite{SCHWINGER,UNRUH,ROVELLI,GHOSE,DU,ALTER1,ALTER2,SAM}. 
Although there has been an attempt to clarify some of these
misunderstandings by the original authors themselves \cite{AAV2}, many
points remain to be clarified. In this section we review the various
criticisms and assess their relevance to the issue of protective
measurements.

\subsection{Are we measuring at all?}

Schwinger \cite{SCHWINGER} raised the following objections to the AAV
proposal:  i) even in the conventional Stern-Gerlach set up, as the
SG-field is weakened, the two beams begin to overlap and no
SG-measurement is performed, ii) repeated SG-measurements have already
demonstrated the probability amplitude (epistemological) interpretation
of the wave function.

Unlike the response of Aharonov and Anandan to this \cite{science}, we
do agree with Schwinger that the effective SG-field is weak, because of
the ${1\over T}$ factor. But the circumstances are otherwise quite
different from an usual SG-measurement. Since the interaction time in
protective measurements is very large, even a weak SG-field is able to
produce a measurable shift in the apparatus pointer position.

Regarding the second point made by Schwinger, it should be emphasized
that AAV do not claim to associate reality with all wave functions. For
example, the wave function for unstable systems can only be interpreted
statistically. Also, repeated modified-SG (protective) measurements are
indeed consistent with treating the wave function as `` real".

\subsection{Are we measuring a known state?}

Rovelli\cite{ROVELLI}, and, Samuel and Nityananda\cite{SAM} have objected
to this proposal on the ground that the fact that the wave function
does not collapse is a trivial consequence of it being an
eigenstate of the dominant Hamiltonian to start with.  Though what they
say about entanglement is correct, they overlook the crucial fact that
the shift in the pointer is proportional to the expectation value of an
operator which {\em doesn't commute} with this dominant Hamiltonian. Thus one
{\em measures} the expectation value of an arbitrary operator of the
system, while the wave function doesn't collapse for obvious reasons.

Another objection of these authors is that the wave function has to be
known {\it a priori} in order to make a protective measurement.  This
claim is not completely correct, because all that is required in the
analysis of protective measurements is that the system is in a
non-degenerate eigenstate of its Hamiltonian, allowing for the
possibility of the situation where the Hamiltonian and the state may be
unknown. Indeed, one can find situations where one may know that a
system is in an eigenstate without knowing the Hamiltonian. An example
is a trapped atom, where the potential may not be known before hand, but
one does know that after a sufficiently long time the atom is to be
found in the ground state. Protective measurement, in principle, allows
the measurement of any operator of the trapped particle, without
destroying the state.

Alter and Yamamoto \cite{ALTER1} have constructed an interesting
example of a type of measurement whereby the system (called signal by
them)  and the apparatus (called the probe)  maintain {\em exact}
disentanglement after the measurement. This is achieved by using the
following interesting property of coherent states of a harmonic
oscillator: For a Hamiltonian 
$\hat H=\hbar\kappa({\hat s}^{\dag}\hat p+\hat s{\hat p}^{\dag})$,

\beeq
\hat U(t) |\beta\rangle_s |\gamma\rangle_p=
|a \beta-ib\gamma\rangle_s |a\gamma-ib\beta\rangle_p ,
\eneq
where $\hat U = e^{i{\hat H}t}$, and $\hat s,{\hat s}^{\dag},\hat p,{\hat p}^{\dag}$ are the annihilation
and creation operators of the system and probe respectively;
further, $a=\cos {\kappa t}$ and $b=\sin {\kappa t}$.  Now they take the
squeezed coherent state $|\alpha,r\rangle_s$ as the system state and
the squeezed vacuum state $|0,q\rangle_p$ as the probe state. The
abovementioned property of coherent states then implies that the
disentangled state $|\alpha,r\rangle_s |0,q\rangle_p$ remains
disentangled under the unitary evolution $\hat U$ provided $q=-r+i\phi$
for any arbitrary phase $\phi$. Their idea is then to make a
measurement on the probe to infer an observable in the signal
state, undo the ``deterministic change" of the system by driving it back
to its original state through a classical field, and repeat this process
as many times as one needs. They have called this a ``protective
measurement" because measurements are being carried out on the system
while maintaining the ability to restore the system to its original
state. The price they had to pay for this was the full a priori
knowledge of the system state.  Hence they concluded that full a priori
knowledge of the state is needed for protective measurements.

Aharonov and Vaidman \cite{AV1} criticized this work on the basis that
the squeezed state they use is not a non-degenerate eigenstate of the
harmonic oscillator Hamiltonian, and hence does not satisfy the
criterion for protective measurement. Also, \cite{AV1} claim that the
scheme of Alter and Yamamoto allows for disentanglement to be
maintained only when certain observables are measured, much the same
way as in eigenstate measurement or in ``ideal von Neumann"
measurements. In their rebuttal to this, Alter and Yamamoto
\cite{ALTER2} have emphasized that one can measure {\em all} the
observables associated with the signal. They further asserted that in
their scheme entanglement is {\em exactly} avoided while the protective
measurement scheme of AAV avoids this only approximately.  We fully
agree with this latter remark, and shall analyze its true import a
little later.

As we see it, the scheme of \cite{ALTER1} is quite different from that
of AAV and suffers from the requirement of full a priori knowledge of
the state which is not a restriction on the AAV proposal. On the other
hand this scheme is attractive because it avoids
entanglement exactly, and is yet another candidate scheme to measure
expectation values of observables in the single quantum state without
irretrievably destroying it. To this extent it appears reasonable to
call the scheme of \cite{ALTER1} also a protective measurement, even if
the single quantum state does not satisfy the criterion laid out by
AAV.

One of the objections raised by Ghose and Home \cite{GHOSE}  (in
addition to stating that protective measurements require the
specification of the state) is that Aharonov {\it et al} have not
solved the problem of wave function collapse.
Protective measurement does not solve
the problem of wave function collapse, and Aharonov {\it et al} have
not claimed otherwise as they state quite explicitly in \cite{AAV2}.
The crucial point here is that there is no entanglement between the
system and the apparatus after the adiabatic interaction. So, if an
actual measurement, by whatever mechanism, is made on the apparatus,
which {\em irreversibly} registers the outcome, the wave function of
the {\em system} will not collapse. This is similar to an eigenstate
measurement using conventional method, where the wave function of the
system does not change during the process of measurement, so the
question of collapse, as far as the wave function of the system is
concerned, doesn't arise.  The wave function of the {\em apparatus}, on
the other hand, does ``collapse'' in the sense that the outcome has to
be registered in an irreversible way. This aspect of the measurement
problem is certainly not solved by protective measurements.

\subsection{Is the final state entangled?}

The most serious attack on the idea of protective measurements can
be made on the ground that in realistic situations, the wave function of
the system apparatus combine is still entangled, though the degree of
entanglement can be made arbitrarily small, the probability of finding the 
system in a state orthogonal to the initial state being of order 
${1\over T^2}$. 
This is so because in first order perturbation theory the correction to the 
energy eigenstate is orthogonal to it.
For ensemble measurement,
this small ``corruption'' is inconsequential as it will affect the
distribution of  the outcome very little. By working with suitably large
ensembles one can isolate and control this admixture. This is the reason 
why adiabatic theorem works in the conventional interpretation of
quantum mechanics. For a single system, however,
even an extremely tiny entanglement can have disastrous consequence
as a single measurement can yield any outcome whose probability 
is non-zero, resulting in a collapse to the small
admixture. 

The issue of entanglement has also been raised by
Choudhury, Dasgupta and Datta \cite{DU} as well as Alter and Yamamoto
\cite{ALTER2}.  
We have, however, some objections to the technical
treatment of Ref. \cite{DU}. They use small time
evolution equations repeatedly in their paper,  
make unwarranted restrictions like  simultaneous commutativity (or lack of it) 
of $Q_A$, $Q_S$
with $H_A$, $H_S$ respectively, etc. They also argue, fallaciously,
that entanglement persists even in the adiabatic limit. This is a
consequence of their ignoring the fact that the support for the
wavefunction where this happens is exponentially small.

However, these authors stress the point that there
are subtleties regarding the reading of the pointer position. 
In fact they correctly emphasise the point that the spread in the 
wavepacket
of the apparatus must be handled and that the burden of protective measurements
is passed on to a measurement of the pointer position.
We have
fully analyzed this problem in sections IV and V.

We fully concur with Alter and Yamamoto \cite{ALTER2} regarding the
serious consequences of entanglement, however small, for measurements
on single systems. As a practical remedy, one could use a small number
of systems prepared in identical states so that the small entanglement
would not spoil each of the protective measurements performed on
this small number. That, however, precludes attaching any ontological 
meaning to the wave function. 

\section{``Reading out'' the Pointer Position}

\subsection{``Spreading'' of the pointer}

Having established the fact that an adiabatic interaction makes it
possible that the center of the wave packet of the pointer shifts by an
amount proportional to the expectation value of the measured observable,
we now move over to the issue of retrieving the information about the
center of the wave packet. One can see that in any setup for protective
measurements the pointer wave packet will spread simply because the
detected pointer variable doesn't commute with the free Hamiltonian of the
apparatus. Condition for adiabaticity requires that the interaction of the
system with the apparatus be for as long a duration as possible. However,
the increased spreading of the wave packet of the pointer would interfere
with resolving the shift of the center.
This aspect of protective measurements was completely overlooked
in the original AAV proposal and, as we shall see in this section, it
is crucial for protective measurements to work.

In order to obtain a detectable shift in the pointer position, it seems
reasonable that the increase in the width of the wave packet should be
at least smaller than the shift.  In the example discussed in section
IID, we may compare the square of the width of the wave packet
($\epsilon^2+{T^2\over M^2\epsilon^2}$) with the square of the shift in
the position of the wave packet, which is $\langle Q_S\rangle_\nu$.
Thus, to have a good measurement, $T < {\langle Q_S\rangle_\nu\epsilon M}$.
From this expression one can see that in order to increase $T$, as
one would desire for an adiabatic interaction, one can only increase
the mass $M$ of the particle. On the other hand, if the measured
expectation value $\langle Q_S\rangle_\nu$ is very small, $T$ also has
to be small in order to resolve the shift in the pointer from the
spread.  So even in the case $[Q_A,H_A]=0$, the spreading of the
wave packet is unavoidable, and hence puts a limit on the time of the
interaction, which in turn would interfere with making the interaction
adiabatic.

From the analysis of the case $[Q_A,H_A]=0$, one would recall that the
initial apparatus state is a wave packet of eigenstate of the operator
conjugate to $Q_A$. Now because $[Q_A,H_A]=0$, that operator doesn't
commute with $H_A$. This will lead to a spreading of the wave packet under
the action of the free Hamiltonian of the apparatus $H_A$. In order that
the wave packet doesn't spread very fast, the initial width of the wave
packets should not be too small. The spread will be more as time increases,
and so one should try to keep the measurement time as small as possible to
avoid spreading. But in protective measurements the interaction has to
be adiabatic. 
So, one has to strike a balance between the
spreading of the wave packet and the time of interaction. 

Several conceptual issues arise even though the general formalism shows
a way of measuring expectation values of observables without disturbing
the (single) state. What has been shown is that this protective way of
measurement shifts the pointer position by an amount depending on the
expectation values of observables in the state of the single system as
opposed to being shifted by all possible eigenvalues of the observable
in the conventional measurement picture. The implication is that the
measurement of the pointer position results in a measurement of the
expectation value.

\subsection{Nature of the apparatus}

This raises some fundamental issues. According to the quantum mechanical
lore, no single measurement of an observable in a quantum state yields 
the value of the observable. Among the many critics of the AAV proposal
only Choudhury, Dasgupta and Datta \cite{DU} emphasized this fundamental
problem.
To understand this issue properly
it should be understood that the wave packet (in the example of section II)
was used to model an apparatus. According to the conventional
interpretation of quantum mechanics the apparatus has to be treated as
being ``classical''.  More precisely, the ideal apparatus must satisfy
the following conditions:i)superposition of pointer states should not
be realizable and ii)the outcome of the {\it measurement} of the
pointer state should itself be dispersion-free. That the wave packet
model for the apparatus used had associated with it the dispersion
$\epsilon$ would then be interpreted as an artifact of the model. To
rephrase Penrose \cite {PEN}, {\em even though the model of the apparatus
has not been delicately organized in such a way that the adiabatic
interaction is magnified to a classically observable event, one must
consider that it could have been so organized}. 
Only a more satisfactory model of the apparatus would lead
to a resolution of these issues. It should be stressed that the
requirement of the non-realizability of the superposition of pointer
states is an important prerequisite for any such model and this may
necessitate a more complete analysis including agencies for decoherence
as considered in \cite {ANU}.
If one accepts this
interpretation, a single protective measurement would yield the
expectation value of a chosen observable in the state of the single
quantum system, which moreover, is left undisturbed by the measurement
process. 

The sceptic may argue that when such a consistent treatment of the
apparatus is made, the conclusions of the present analysis may also not hold!
Then one will have to reckon with the quantum nature of the apparatus used
in the foregoing analysis, and introduce the inevitable classical apparatus
at a later stage.

In that case the wave packet dispersion $\epsilon$ should be taken
seriously and  a number of difficulties seem to arise. A single
measurement done on the wave packet will not yield the location of the
center. One possibility is that we consider adiabatic coupling of a
single quantum system to an ensemble of apparatuses and make
measurements on the ensemble of apparatuses to determine the pointer
position.  This is not such an unreasonable arrangement. For
example, the ensemble of apparatuses could be a beam of atoms
interacting adiabatically with the spin of the system.  Such an
ensemble approach inevitably carries with it uncertainty in the
knowledge of the position of the apparatus. However, the pointer
position which is the average of the outcome of these position
measurements, can be determined with arbitrary accuracy.

\subsection{Repeated measurement of a single state}

The reason one was forced to consider an ensemble of
measurements in the conventional measurement was that the (impulsive)
coupling of the system to the apparatus resulted in an entangled
superposition where all possible pointer positions could be realized
with appropriate probabilities. In contrast, in the protective
measurements only a single pointer position is chosen. 
This affords a more
interesting alternative to considering an ensemble of apparatuses as 
argued above. Since the state of the system is unaltered and
the expectation value of observables in the state of the single quantum
system is given by the {\it shift} of the pointer position and not the
pointer position itself, it is possible to consider the coupling between
a single apparatus and the system and make repeated measurements on the
(single) apparatus. Again, the reason why conventional measurements fail
in this regard is that there every act of measurement irretrievably
changes both the system state and the apparatus state. In the case of
the protective measurements too, the state of the apparatus itself is
continually being altered by the measurement in an unpredictable manner.
But the shift between two successive measurements
constitutes a measurement of $\langle Q_S\rangle$ and its average
value can be determined by performing a large number of such
measurements. In practice, the measurement of the position of the
pointer can be made with a suitably small uncertainty and the
subsequent measurement done after an interval not too long to increase
$\epsilon(t)$ but long enough to justify the adiabaticity. Such
considerations will play an important role in practical
implementations.

One must however point out some caveats. Strictly speaking, even if the
wave-packet is sharply peaked, the first measurement of the position
can yield any value not necessarily centered around the mean value.
Whether this will render useless the idea of repeated measurements on a
single apparatus is to be settled by more careful examinations of the
points raised.  This brings us again to the point mentioned earlier
that the wave-packet as a model of the apparatus must provide, if not
dispersion-free measurements, that at least the measured values of the
pointer position are close to its mean.

\subsection{Quantum Nondemolition Measurement of the Apparatus}

There is yet another interesting way out of the problem of measuring
the shift of the wave packet of the pointer. This is based on repeated
weak quantum nondemolition (QND) measurements \cite{BRAGINSKY}
performed on the {\em apparatus}.
Recently Alter and Yamamoto \cite{ALTER3} analyzed the
problem of a series of repeated weak QND measurement on a quantum
system, to address the question of getting information about the 
unknown wave
function of a single quantum system from such measurements. They
concluded that it is possible to obtain the mean value of an observable
in an unknown state, but no information can be obtained about the
uncertainty of the observable. Hence one cannot obtain any information
about the wave function. Also, the state is completely altered in the
process.

Their scheme is best illustrated through the first of the two examples
they consider in \cite{ALTER3}. This is a series of photon number QND
measurements performed on a single wave packet of light. The
probe(apparatus) is a squeezed coherent state $|\alpha_0,r\rangle$ with
real squeezing parameter $r$. The signal and probe are correlated
through an unitary transformation $\hat U=e^{i\mu{\hat n}_s{\hat n}_p}$
, where ${\hat n}_s,{\hat n}_p$ are the photon number operators for the
system and probe respectively. The signal photon number is inferred
from measuring the second quadrature of the probe. A series of such
measurements yields ${\tilde n}_1,{\tilde n}_2,{\tilde n}_3,...$ for
the inferred photon number of the signal.  The photon number
distribution in the unknown initial state is taken to be
$P_0(n)=N[n,n_0,\Delta_0^2]$, with unknown $n_0,\Delta_0^2$, where
$N[x,x_0,\sigma^2]=(2\pi\sigma^2)^{-1/2} exp [-(x-x_0)^2/2\sigma^2]$ is
a normalized normal distribution; here $\Delta_m^2$ is the uncertainty
due to measurements and is controllable as in classical measurements.

With each measurement, the system state {\em changes} and the photon
number distribution of the signal after k measurements becomes
$P_k(n)=N[n,n_0^{(k)}, \Delta_k^2]$, with $n_0^{(k)}=\Delta_k^2[{n_0
\over \Delta_0^2}+{\sum{\tilde n}_i\over \Delta_m^2}]$ and
$\Delta_k^2=({1\over \Delta_0^2}+{k\over \Delta_m^2})^{-1}$.  The
important features of this example to concentrate on are: i)
$P_k(n_0^{(k)})$, the diffusion of the centre after k measurements, is
given by $N[n_0^{(k)},n_0, (k/\Delta_m^2)\Delta_0^2\Delta_k^2]$.{\em
This distribution is centred at $n_0$}.  ii)If $\bar
n=\sum_{i=1}^k{\tilde n}_i/k$ and $\bar{\Delta n^2}=\sum_1^k({\tilde
n}_i-\bar n)^2/(k-1)$ are the mean and variance of the outcome of
measurements ${\tilde n}_1,{\tilde n}_2,...$, the probability
distribution of $\bar n$ and $S=[(k-1)/\Delta_m^2]\bar {\Delta n^2}$
are given by $N[\bar n,n_0,\Delta_0^2+\Delta_m^2/k]$ and
$\chi^2[S,(k-1)]$ respectively, where $\chi^2[x,\nu]$ is the
chi-squared distribution of the variable $x$ which is centred at
$\nu$.{\em Thus, while $\bar n$ is a ``good" estimator for $n_0$, $\bar
{\Delta n^2}$ being centered at $\Delta_m^2$ has nothing to do with the
initial uncertainty $\Delta_0^2$}. iii) Eventually, the width of the
distribution $P_k(n)$ becomes zero which means the signal becomes an
eigenstate of photon number with eigenvalue $n_0$.

While the conclusions of \cite{ALTER3} were negative as far as using
repeated weak QND measurements to determine the unknown wave function
of a single system, it appears tailor-made to solve the problem of
``reading the pointer position" in protective measurements. Thus we
apply their scheme not to the system part of the protective measurement
set ups, but to the apparatus part instead. Then we can get information
about the center of the wave packet, which in the protective
measurement scheme carries information on the expectation values of
observables in the system state, through repeated measurement of the
(quantum) apparatus. There is also the added advantage that the
variance in the outcome of these repeated measurements has nothing to
do with the spread in the wave packet of the apparatus. The uncertainty
in the measured values of $\langle Q_S \rangle$ will therefore be more
like errors in classical measurements which are controllable. Hence
there need not be any uncontrolled uncertainty in the reconstruction of
the original state.  The concerns expressed in section IVA get mitigated in
an elegant manner.

This example comes closest to realizing the ideals of a classical
apparatus but nevertheless dealing with an apparatus that is treated
quantum mechanically.  We need not care about the fact that it doesn't
give us information about the variance, as all we need to know, in
order to complete the protective measurement, is the position of the
center of a pointer wave packet. There is an added bonus to this method
in the sense that $n_0$ can be determined as the average of the outcome
of sequence of measurements {\em as well as by performing an eigenstate
measurement on the eventual apparatus state!}.  We also don't care that
the original state is destroyed after the measurement, because for us
it is the state of the apparatus that is destroyed, and not of the
system.

Thus one may proceed with a protective measurement by first
allowing an adiabatic interaction of the system with an apparatus which
can be treated quantum mechanically. This would result in a shifted
wave packet of the pointer. One can then do a series of {\em weak} QND
measurements on this wave packet to get the position of the center.
This seems the most promising possibility for experimentally realizing
protective measurements.

\section{Some final remarks}

\subsection{Restrictions on the apparatus}

From our general discussion of protective measurements, it is clear
that many restrictions may have to be imposed on the kind of apparatus
to be used. By an apparatus in this
context, we shall mean a specification of $H_A,Q_A$.  In the general
case where $[Q_A,H_A]\neq 0$, it is not clear whether the operators $X$
and $Y$ can be physically realized in an actual set up. Also, as
already pointed out, it may not always be possible to even find a $X$
that is canonically conjugate to $Y$.

The other important restriction on the apparatus comes from the
requirement that $|\nu\rangle|a\rangle$ in our general treatment should
not be a degenerate eigenstate of $H_0=H_S+H_A$ unless the perturbation
$gQ_AQ_S$ is {\em diagonal} in the degenerate subspace.
Generically, $H_A$ should not have a continuous spectrum, though in a
specific example given in IID, continuity of the spectrum was not a
problem because the perturbation there was diagonal in the degenerate
subspace. In fact, in all cases where $[H_A,Q_A]=0$, the perturbation
will be diagonal in the degenerate subspace (recall that $|\nu\rangle$
is a {\em non-degenerate} eigenstate of $H_S$). These considerations
rule out, for example, the prototypical Hamiltonian in discussions of
the Stern-Gerlach set up i.e $H=P^2/2M+\mu B_0 \sigma_z+\mu
B_iX\vec\sigma\cdot\vec n$. As emphasized before, AAV in their spin-1/2
example chose $H_A=0$. But once this is relaxed, the difficulties
stressed here become relevant.

The other important point to emphasize is that ${Q_AQ_S\over T}$ should
be a well-defined perturbation over $H_0$ in the sense that its matrix
elements in the basis spanned by the eigenstates of the unperturbed
Hamiltonian should exist. This too rules out the prototypical
Hamiltonian in the discussions of the Stern-Gerlach model mentioned
above, because expectation value of $x$ in any plane-wave state does not
exist. Not only should the matrix elements exist, at least some of the
diagonal matrix elements of $Q_A$ should be nonvanishing, as otherwise
there will be no shift in the pointer position.  This, for example,
rules out a linear position coupling in the case of a Harmonic
oscillator.

One might have have thought that the Stern-Gerlach Hamiltonian could
have been used with some sort of ``regularization" like putting the
particle in a box, or treat the free particle as a harmonic oscillator
with a very tiny $\omega$.  But both these are unsatisfactory for the
purpose of protective measurements because in the first case
$\langle|x|\rangle$ in the eigenstates of $P^2/2M$ with box-boundary
conditions is always the same and is at the center of the box.  Then
the operator $Y$ of our general treatment is the identity operator for
which there is no canonical conjugate. Physically, this means that the
adiabatic interaction only produces an overall phase which is of no
consequence in shifting the pointer position. The second alternative of
treating the free particle as the limit of a harmonic oscillator with
vanishing frequency is also no good as in this case the expectation
value of $x$ in the oscillator energy eigenstates vanishes and there
will be no pointer position shift. In these cases a more rigorous
handling than what perturbation theory offers may be needed.

It is not clear that even the case where $[Q_A,H_A]=0$ is easily
realizable experimentally.  In the example of IID one needs a
momentum dependent magnetic field. While it is always possible to
experimentally create a position dependent magnetic field by using
inhomogeneous fields, it is not clear how one would create the former.

These restrictions on $Q_A,Q_S$ are not warranted in the conventional i.e
impulsive measurements as there $Q_AQ_S$ is dominant and $\int H_A,\int
H_S$ can be neglected in comparison (the integration is over the
duration of the impulse).

\subsection{Does it {\it really} work for a single system?}

In the entire analysis it has been assumed that entanglement effects
(between the apparatus and the system) can be made arbitrarily small as
$T$ is made large. In the case of conventional measurements, a small
contamination of the wave function will also have only a small
statistical effect. With a large enough ensemble of states, the effect
of such small admixtures in the wave function can be controlled. In the
case of protective measurements the situation is radically
different. However small the amplitude for entanglement in the large $T$
limit, the outcome of the first measurement on the single system can
always  be states of the system and apparatus which are part of the
small amplitude. This would have a deleterious effect on the subsequent
measurements. It is clear that this potential problem persists no matter
how large $T$(or how small ${1\over T}$) is made. 
Stated differently, however large T is made, the {\it possibility} that
even a "protective measurement" projects the system into a state orthogonal
to its initial state can never be ruled out. The fact that the {\it calculated
probability } for this to happen could be extraordinarily tiny is of no
consequence because for a single system under such circumstances, probabilistic
concepts are inapplicable. To illustrate this in the specific context of the
example of sec IIE, the angle $\theta$ is always non-zero though very small
and the original spin is precessing around the unknown magnetic field with this
inclination. Quantum mechanically speaking, any measurement can realise both
the initial state $|\tilde +>$ as well as its orthogonal complement 
$|\tilde ->$.
This may well be the
most formidable obstacle to realizing protective measurements with certainty.  {\em In
this sense it is the conventional interpretation of the wave function
and measurements that is protected against the vagaries of statistical
fluctuations}

\subsection{Philosophical issues}

The idea of  a protective measurement, like its conventional
counterpart, also has some philosophical issues associated with it.
Because of the fact that there exists a possibility of measuring the
expectation value of an observable from an unknown wave function when
it is an energy eigenstate, one might tend to associate a reality with
energy eigenstates.  If one believes what can be measured is ``real'',
then the energy eigenstates appear, on first glance, to satisfy this
condition, and seem to be special in this regard.  On the other hand,
the ubiquitous tiny entanglement makes it impossible to make a
measurement on one single system, with complete certainty, as we
discussed before.  If the entanglement is really tiny, it may not be
that bad from a practical point of view, in the sense that a small
number of such measurements are likely to give the right answer.  But
it still precludes associating a ``reality''  with the wave function of
a {\em single} system.

Unruh has raised an objection to associating a ``reality'' with the
wave function even after assuming the validity of the idea of
protective measurements on a single system \cite{UNRUH}.  He argues
that the energy eigenstates may be considered to have a ``reality'',
but that cannot be concluded about any arbitrary state.  For the
reasons mentioned above, we do not believe one should associate a
``reality'' even with the energy eigenstates of a system.

Unruh also points out that ``protection'' in the sense used by AAV is 
an attribute which a
system either already has or doesn't have, which means that only if a
system is already in a non-degenerate energy eigenstate, can a
protective measurement be performed on it. One cannot ``protect'' a
given unknown wave function.

\section{Summary}

In summary, we have critically examined the idea of protective
measurement of a quantum state. We have shown that the idea can be
generalized to the case where the interaction Hamiltonian does not
commute with the free Hamiltonian.  We have also looked at earlier
criticisms of the idea and conclude that most of them are not relevant
to the original proposal. 
The relevant criticisms, we believe, are the comments by Alter and 
Yamamoto\cite{ALTER2} on the omnipresent infinitesimal entanglement, comments 
by Choudhury, Dasgupta and Datta\cite{DU} pointing out the subtleties in 
reading out the pointer, and the comments by Unruh\cite{UNRUH} on the 
interpretation of protective measurements. 
We discuss various conceptual issues involved
in the process of protective measurements and infer that there are
several constraints imposed on the measuring apparatus.  It is pointed
out that a single measurement does not yield any information. We
propose two schemes as a way out of this problem. One of this involves
performing repeated measurements on the single quantum system, making
use of the fact that the system wave function doesn't change. The other
proposal involves performing a series of quantum nondemolition
measurements on the {\em apparatus}, which is to be treated quantum
mechanically, after a single protective measurement on the quantum system.
After analyzing all the issues involved, we conclude that
although experimentally realizing protective measurements is a
possibility,  one can never perform a protective measurement on a
single quantum system with absolute certainty because of the tiny
unavoidable entanglement which is always there.  
This is sufficient ground for precluding the ``reality''
of the wave function. 
In this sense we agree with Unruh that what the
AAV proposal has achieved is a fresh understanding of the nature of
measurements in quantum mechanics, rather than elevate the wave function
to a new status.

On the practical side, it appears that protective measurements  
(where possible)  can be
used to determine the wave function using considerably smaller ensembles
than in traditional measurements, with the added bonus that the ensemble 
is practically left in tact after the measurements.
We give the following semi-quantitative argument in support of this.
Of course, a more detailed model-specific analysis would be required to make
these arguments more concrete.

Let us compare the measurement of some quantity $X$ in both conventional
ensemble measurements as well as in protective measurements. In the latter
case, let us consider doing it with an ensemble of $N_p$ identically prepared
states and let the
former be done with an ensemble of $N_c$ identically prepared states. In the
case of protective measurements we obtain with probability $1-c^2/T^2$,
the
exact expectation value $\langle X\rangle _{exact}$ and with probability $c^2/T^2$ the
expectation value $\langle X\rangle _{\perp}$ where $c$ depends on the details of the system
and $\perp$ refers to the normalised state in the subspace normal to the 
initial state as picked out by first order perturbation theory. It is worth
noting that the relative probability $\sim 1/T^2$ as in first order 
perturbation the change in the wavefunction is {\it orthogonal} to it.
This
works to a tremendous advantage for realising protective measurements.
Thus
the error in the estimation of the expectation value by protective measurements
is $c^2\langle X\rangle _{\perp}/T^2$. Of course the statistical error $1/\sqrt{N_p}$ weighed
with the relevant probability should also be taken into account. Combining the
errors in quadrature, one gets for the estimate of error in protective
measurements
\begin{equation}
\epsilon_p = {c^2\over T^2}\sqrt {[\langle X\rangle _{\perp}^2+{1\over N_p}]}
\end{equation}
The size of the conventional ensemble $N_c$ required to match this precision
is roughly $1/\epsilon_p^2$ and is given by
\begin{equation}
N_c = {T^4\over c^4}{1\over [ \langle X\rangle _{\perp}^2+1/N_p]}
\end{equation}
Thus with large enough T one can achieve substantial reduction in the ensembles
required for protective measurements, for any given degree of precision in
measurements.The estimate provided above is crucially dependent on ones ability
to carry out the QND measurement on the apparatus as detailed in sec IVD. 

This is indeed a very attractive practical spin-off for the AAV proposal.
The
other attractive feature, as already mentioned earlier, is that the original
pure ensemble remains pure with probability $1-c^2/T^2$ whereas in
conventional measurements the original pure ensemble is {\it completely}
destroyed in the sense that it is reduced to a mixed ensemble from which it can not be reconstructed.

\acknowledgments
One of us (TQ) thanks the Institute of Mathematical Sciences for making
possible a stay, during which this work was completed, and providing
warm hospitality.

\end{document}